\documentclass[prl,aps, twocolumn, amsmath,amssymb,superscriptaddress]{revtex4-1}
\usepackage[dvips]{epsfig}
\usepackage{float}
\usepackage{amsmath}
\usepackage{amssymb}
\usepackage{amsfonts}
\usepackage{euscript}
\usepackage{enumerate}
\usepackage{hhline}
\usepackage{threeparttable}
\usepackage{pslatex}
\usepackage{tabularx}
\usepackage{color}
\usepackage{IEEEtrantools}
\usepackage{graphicx}% Include figure files
\usepackage{dcolumn}% Align table columns on decimal point
\usepackage{bm}% bold math
\usepackage[sort&compress]{natbib}

\makeatletter
\renewcommand{\@biblabel}[1]{#1. }
\renewcommand{\@dotsep}{500}
\renewcommand{\@pnumwidth}{0em}
\renewcommand{\l@figure}[2]{% #1 is e.g. Figure 1 + caption, #2 is pg.
\@dottedtocline{1}{1.5em}{2em}{Figure #1}{}\vspace{15pt}}

\begin{document}

\preprint{Draft 4}

\title{Improving the performance of bright quantum dot single photon sources using amplitude modulation}% Force line breaks with \\

\author{Serkan Ates}\email{serkan.ates@nist.gov}
\affiliation{Center for Nanoscale Science and Technology, National
Institute of Standards and Technology, Gaithersburg, MD 20899,
USA}\affiliation{Maryland NanoCenter, University of Maryland,
College Park, MD 20742, USA}
\author{Imad Agha}
\affiliation{Center for Nanoscale Science and Technology, National
Institute of Standards and Technology, Gaithersburg, MD 20899,
USA}\affiliation{Maryland NanoCenter, University of Maryland,
College Park, MD 20742, USA}
\author{Angelo Gulinatti}
\affiliation{Politecnico di Milano, Dipartimento di Elettronica e
Informazione, Piazza da Vinci 32, 20133 Milano, Italy}
\author{Ivan Rech}
\affiliation{Politecnico di Milano, Dipartimento di Elettronica e
Informazione, Piazza da Vinci 32, 20133 Milano, Italy}
\author{Antonio Badolato}
\affiliation{Department of Physics and Astronomy, University of
Rochester, Rochester, NY 14627, USA}
\author{Kartik Srinivasan} \email{kartik.srinivasan@nist.gov}
\affiliation{Center for Nanoscale Science and Technology, National
Institute of Standards and Technology, Gaithersburg, MD 20899, USA}

\date{\today}% It is always \today, today,

\begin{abstract}
Single epitaxially-grown semiconductor quantum dots have great
potential as single photon sources for photonic quantum
technologies, though in practice devices often exhibit non-ideal
behavior. Here, we demonstrate that amplitude modulation can improve
the performance of quantum-dot-based sources. Starting with a bright
source consisting of a single quantum dot in a fiber-coupled
microdisk cavity, we use synchronized amplitude modulation to
temporally filter the emitted light. We observe that the single
photon purity, temporal overlap between successive emission events,
and indistinguishability can be greatly improved with this
technique. As this method can be applied to any triggered single
photon source, independent of geometry and after device fabrication,
it is a flexible approach to improve the performance of solid-state
systems, which often suffer from excess dephasing and multi-photon
background emission.
\end{abstract}

 \maketitle

Solid-state quantum emitters are potentially bright, stable, and
monolithic sources of triggered single photons for scalable photonic
quantum information
technology~\cite{ref:Lounis_SPS,ref:Santori_Book}. Source properties
which must be optimized for applications include the fraction of
photons emitted into a useful optical channel, the repetition rate
at which the source is operated, the degree to which multi-photon
emission is suppressed, and the extent to which the single photons
are identical. One specific solid-state system that has drawn
considerable interest is the InAs/GaAs quantum dot (QD)
heterostructure. Despite significant development of these sources,
achieving good performance with respect to all of the aforementioned
parameters can be
challenging~\cite{ref:Shields_NPhot,ref:Michler_book_2009}. For
example, the high refractive index contrast between GaAs and air
requires modification of the geometry to prevent most of the QD
emission from remaining trapped within the semiconductor. The
existence of radiative states within the QD heterostructure that are
spectrally resonant with the transition of interest can limit the
single photon purity of the emission. Interactions between the
excitonic transition and electronic carriers and phonons in the host
semiconductor can cause dephasing that prevents the emitted photons
from being perfectly indistinguishable.

Researchers have developed a number of tools to address these
limitations.  Nanofabricated photonic structures can ensure that a
significant fraction of the QD emission is
collected~\cite{ref:Pelton,ref:Solomon,ref:Strauf_NPhot,ref:Claudon,ref:Davanco_BE}.
Optical excitation resonant with excited states of the QD can limit
multi-photon emission~\cite{ref:Santori_NJP}, increase the coherence
time, and improve the degree of
indistinguishability~\cite{ref:Santori2,ref:Ates_PRL09,ref:Weiler_Michler_PSS_B}.
Purcell-enhancement of the radiative rate through modification of
the QD's electromagnetic environment~\cite{ref:Gerard1} can also
produce single photon wavepackets that are more
indistinguishable~\cite{ref:Santori2,ref:Varoutsis_PRB05,ref:Weiler_Michler_PSS_B};
furthermore, it increases the maximum repetition rate at which the
source can be operated.

Here, we describe a different approach to improving the performance
of QD single photon sources (SPSs). Rather than influencing the QD
radiative dynamics, we instead use temporal filtering through
electro-optic amplitude modulation to process and purify the QD
emission.  Synchronized modulation of single photon wavepackets has
recently been demonstrated for both
atomic~\cite{ref:Kolchin_PRL_08,ref:Specht_NPhot_09} and QD
systems~\cite{ref:Rakher_APL_2011}, but those works focused
primarily on demonstrating that modulation was possible and the
variety of wavepacket shapes that it could produce. We begin by
demonstrating a bright, fiber-coupled SPS ($>\,20\,\%$ overall
collection efficiency into the fiber) based on a QD in a microdisk
cavity, and then show that the ability to temporally select portions
of the emitted signal can lead to large improvements in the purity
and indistinguishability of the source. In particular, we
demonstrate an improvement in the single photon purity by a factor
as high as $8$, enough temporal separation between successive
emission events to achieve a 0.5\,GHz repetition rate source, and an
improvement in the two-photon wavepacket overlap by a factor of $2$.
In contrast to other approaches which require modification of the
source, this technique can be applied to any existing solid-state
triggered SPS, regardless of the device geometry and excitation
method (optical or electrical), and can thus be a versatile resource
when implementing solid-state SPSs in quantum information
applications.

\begin{figure*}
\centering \centerline{\includegraphics[width=16
cm,clip=true]{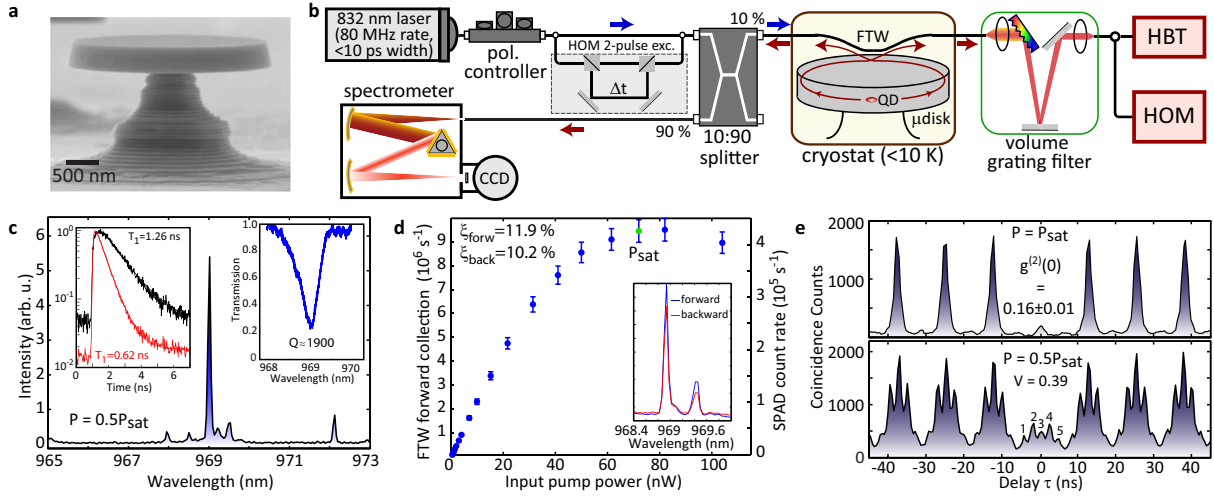}} \caption{\textbf{Bright, fiber-coupled
microcavity-QD single photon source.} \textbf{a}, Scanning electron
microscope image of the GaAs microdisk cavity. \textbf{b},
Experimental setup (details in the Supplementary
Information~[\onlinecite{ref:QD_amplitude_modulation_note}]).  The
10:90 directional coupler allows for simultaneous measurement of the
forward and backward channels of the FTW.  Typically, the emission
spectrum is monitored through the backward channel while the forward
channel is spectrally filtered to select the desired QD transition,
which is then used in subsequent photon correlation measurements.
HBT=Hanbury-Brown and Twiss setup; HOM=Hong-Ou-Mandel
interferometer. \textbf{c}, Photoluminescence (PL) spectrum for a
QD-microdisk device. The left inset shows the PL decay of the QD
line at 969\,nm, both when the cavity is detuned (black) and
on-resonance (red). The right inset shows a transmission spectrum of
the microdisk. \textbf{d}, Spectrally-filtered QD emission as a
function of pump power, where the right $y$-axis shows the detected
count rate on a Si SPAD and the left $y$-axis shows the
corresponding photon count rate collected into the forward direction
of the FTW. $P_{\text{sat}}$ is the pump power at which the QD
emission is highest. The inset shows the measured PL spectrum from
both the forward (blue) and backward (red) direction of the FTW.
\textbf{e}, Upper panel: Second-order correlation function measured
at $P_{\text{sat}}$. Lower panel: Photon indistinguishability
measurement. The suppression of peak 3 with respect to peaks 2 and 4
is due to the two-photon interference effect with
$V\,=\,0.39\,\pm\,0.05$ (See Supplementary
Information~[\onlinecite{ref:QD_amplitude_modulation_note}]). Error
bars in \textbf{d} come from the fluctuation in the detected count
rates, and are one standard deviation values.  The uncertainty in
the $g^{(2)}(0)$ values is given by the standard deviation in the
area of the peaks away from time zero, and leads to the uncertainty
in $V$.} \label{fig:fig1}
\end{figure*}

\vspace{5mm} \noindent \textbf{Efficient fiber-coupled single photon
source}

We use a self-assembled InAs QD embedded in a GaAs microdisk cavity
(Fig.~\ref{fig:fig1}a) as a triggered SPS. Our main objective is to
produce a bright source under pulsed excitation. We use relatively
small diameter ($D\lesssim2.9\,\mu$m) devices to obtain a high QD
spontaneous emission coupling fraction $\beta$ into the resonant
whispering gallery modes (WGMs) of the microdisk. Efficient
outcoupling of the WGMs is achieved using a fiber taper waveguide
(FTW), an approach previously used to create fiber-coupled
microdisk-quantum-dot lasers~\cite{ref:Srinivasan12} and waveguide
SPSs~\cite{ref:Davanco_WG}. Out-coupling of a WGM through the FTW is
quantified by an efficiency $\eta$, whose value is experimentally
determined by measuring the transmission spectrum of the cavity (see
Supplementary
Information~[\onlinecite{ref:QD_amplitude_modulation_note}]). The
overall collection efficiency of photons into each channel of the
FTW is $\xi\,=\,\beta\eta$, in the limit of unity QD radiative
efficiency.

The setup shown in Fig.~\ref{fig:fig1}b is used to measure the
low-temperature micro-photoluminescence spectrum of a microdisk-QD
device shown in Fig.~\ref{fig:fig1}c, where a bright excitonic line
is observed on top of a broad cavity mode at 969\,nm. The relatively
low quality factor mode (Q\,=\,1900; see inset transmission
spectrum) results in a Purcell factor $F_P\,=\,2$, as determined by
measuring the emission lifetime when the QD is on-resonance with the
mode and far-detuned from it (inset to Fig.~\ref{fig:fig1}c). The
brightness of the QD source is determined through the excitation
power-dependent intensity of the filtered signal, which is directly
measured with a Si single-photon avalanche diode (SPAD), as shown in
Fig.~\ref{fig:fig1}d. The right vertical axis is the measured count
rate at the detector, while the left axis is the photon count rate
coupled to the forward channel of the FTW, factoring in the losses
due to spectral filtering, detection efficiency of the SPAD, and the
transmission of the FTW (see Supplementary Information). At
saturation, a collection efficiency $\xi\,=\,11.9\,\%\,\pm\,0.6\,\%$
into the forward channel of the FTW is estimated. Ideally,
collection into the backward channel will equal that into the
forward channel; for this device, we measure a slight reduction (by
14\,$\%$) in the backward channel (inset of Fig.~\ref{fig:fig1}d),
most likely due to asymmetric losses in the setup.  This yields
$\xi\,=\,10.2\,\%\,\pm\,0.6\,\%$ for the backward channel, so that
if both channels are combined, the overall collection efficiency
into the FTW is $\approx\,22\,\%$.

\begin{figure*}
\centering\centerline{\includegraphics[width=0.85\textwidth,clip=true]{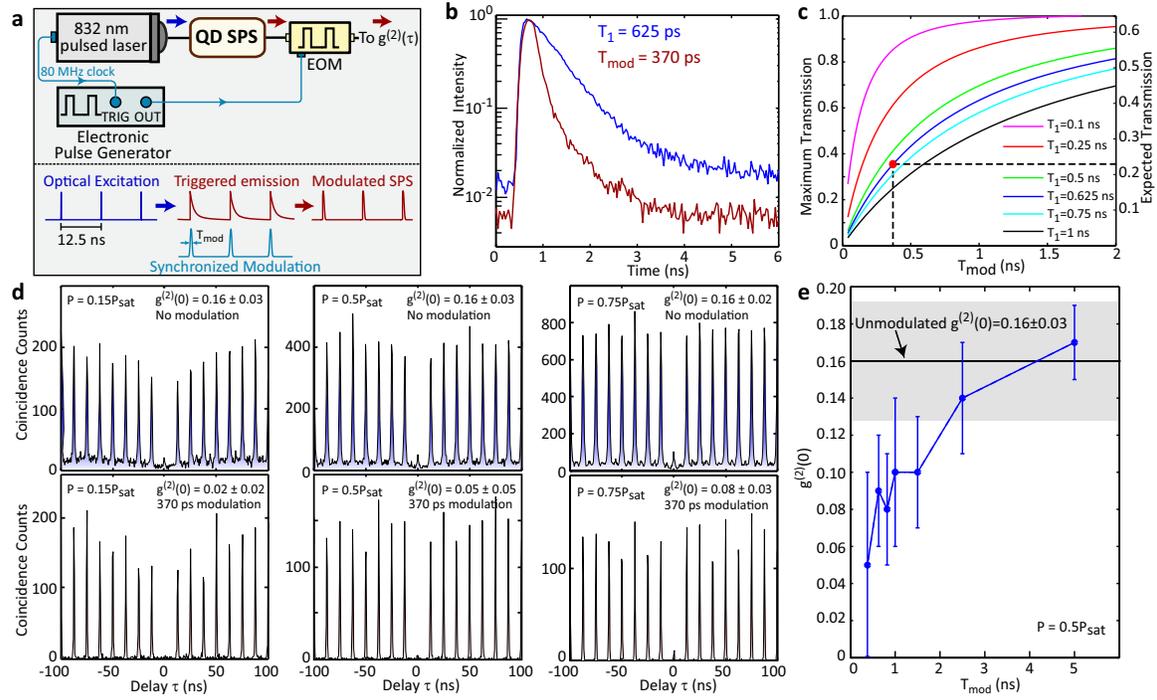}}
\caption{\textbf{Improved single photon purity using amplitude
modulation.} \textbf{a}, Schematic of the amplitude modulation
setup. EOM=electro-optic modulator. \textbf{b}, QD lifetime traces
under no modulation (blue) and 370\,ps modulation (red). \textbf{c},
Calculated transmission through the EOM as a function of modulation
width $T_{\text{mod}}$, for varying $T_{1}$. The left y-axis shows
the maximum possible transmission, while the right y-axis includes
1.9\,dB of insertion loss through the EOM.  The highlighted point is
the expected transmission level assuming $T_{1}$\,=\,625\,ps and for
the 370\,ps modulation used in subsequent experiments. \textbf{d},
Pump-power-dependent second order correlation measurements, without
modulation (top) and with modulation (bottom). \textbf{e},
Modulation-width-dependent second order correlation measurement at
0.5$P_{\text{sat}}$. The shaded gray region corresponds to
$g^{(2)}(0)\,=\,0.16\,\pm\,0.03$, measured for no modulation. The
uncertainty in the $g^{(2)}(0)$ values in \textbf{d} and \textbf{e}
is given by the standard deviation in the area of the peaks away
from time zero.}\label{fig:fig2}

\end{figure*}

The single-photon nature of the collected QD emission is
demonstrated by measuring the second-order photon correlation
function $g^{(2)}(\tau)$ using a Hanbury-Brown and Twiss (HBT)
setup~\cite{ref:QD_amplitude_modulation_note}.  Even at $P_{sat}$,
the pump power for which the emission is highest, we observe (upper
panel of Fig.~\ref{fig:fig1}e) a clear suppression of the
correlation peak at zero time delay, with
$g^{(2)}(0)\,=\,0.16\,\pm\,0.01\,<\,0.5$. We also characterize the
indistinguishability of the single photon emission using a
Hong-Ou-Mandel (HOM)
interferometer~\cite{ref:Hong_Ou_Mandel_PRL,ref:Santori2}, where
consecutively emitted photons are overlapped on a beamsplitter. As
discussed in Ref.~\onlinecite{ref:Santori2} and in the Supplementary
Information, the degree of indistinguishability is experimentally
related to $M\,=\,\frac{A_{3}}{A_{2}+A_{4}}$, where $A_{2,3,4}$ are
the areas of the peaks labeled in the lower panel of
Fig.~\ref{fig:fig1}e. For our QD SPS with
$g^{(2)}(0)\,=\,0.16\,\pm\,0.01$ at $0.5P_{\text{sat}}$
(Fig.~\ref{fig:fig2}(d)), $M\,<\,0.57$ can only occur if there is
two-photon interference (see Supplementary Information).
$M\,=\,0.40$ is observed in Fig.~\ref{fig:fig1}e, indicating a
degree of indistinguishability that is quantified by the two-photon
wavepacket overlap
$V\,=\,0.39\,\pm\,0.05$~\cite{ref:QD_amplitude_modulation_note}.

Since $\beta$ approaches 50$\,\%$ in these devices (half into each
of the clockwise and counterclockwise modes), improving the
brightness of this source requires an increase in $\eta$, which is
estimated to be $\approx$25\,$\%$ in the current
devices~\cite{ref:QD_amplitude_modulation_note}.  This would require
improved overlap between the FTW and microdisk WGMs, through
adjustment of the microdisk and FTW dimensions. While the
demonstrated brightness of $\approx\,11\,\%$ (22\,$\%$) into one
(both) channel of the fiber is smaller than the collection into the
first optic in recent
demonstrations~\cite{ref:Strauf_NPhot,ref:Claudon}, it has the
advantages of being directly fiber-coupled, exhibiting Purcell
enhancement with a relatively low $g^{(2)}(0)$ value at saturation,
and having indistinguishability with a two-photon wavepacket overlap
of 39~$\%$. The direct fiber coupling allows us to easily interface
the bright QD emission with the wide variety of fiber-coupled
optoelectronic components available in the 980\,nm region.  Among
these is a fiber-coupled electro-optic amplitude modulator, which we
use in the following sections to manipulate the purity, repetition
rate, and indistinguishability of the QD SPS.

\vspace{3mm} \noindent \textbf{Improving the purity of the single
photon source}

A non-zero value of $g^{(2)}(0)$ is commonly measured in QD SPSs,
and indicates the presence of temporally coincident multi-photon
emission (within the timing resolution of the system).  Such
emission can originate from other spectrally-resonant radiative
transitions in the system that arise due to the nature of the QD
confinement, which supports a quasi-continuum of (multi)excitonic
transitions whose emission can be enhanced by the presence of a
cavity
mode~\cite{ref:Winger2009,ref:Chauvin_Fiore,ref:Laucht_Finely_2010}.
Another process that can lead to $g^{(2)}(0)\,>\,0$ is carrier
recapture on a time scale comparable to the QD radiative lifetime,
which can enable the emission of more than one photon per excitation
pulse~\cite{ref:Santori_NJP,ref:Peter_APL,ref:Aichele_Zwiller_Benson}.
In this section, we show how amplitude modulation can reduce the
multi-photon contribution, thereby improving the purity of the QD
SPS.

The modified setup is shown in Fig.~\ref{fig:fig2}a.  The trigger
output of the 832\,nm excitation laser is used to synchronize an
electronic pulse generator whose output drives a fiber-coupled,
980\,nm band electro-optic modulator (EOM). Spectrally-filtered QD
emission is fed into the EOM, and its output is sent to the HBT
setup for photon correlation measurements. The pulse generator
produces optical pulses of width
$T_{\text{mod}}>350$\,ps~\cite{ref:QD_amplitude_modulation_note},
measured as the full-width at the 1/e point. The separation between
the EOM gates and the incoming QD emission can be controlled with ps
resolution.

Figure~\ref{fig:fig2}b shows the QD lifetime measured with and
without amplitude modulation, where the modulation produces
$T_{\text{mod}}$\,=\,370\,ps$\,\pm\,20$\,ps, and its extinction
level is $>$20\,dB. Amplitude modulation is expected to reduce the
overall source brightness, both through its temporal gating function
and broadband insertion loss. The transmission through the temporal
gate can be estimated by considering the overlap of the EOM response
and the QD emission~\cite{ref:QD_amplitude_modulation_note}.
Assuming that the EOM gate position is optimal and that the QD
emission follows a decaying exponential, Fig.~\ref{fig:fig2}c shows
the expected transmission level through the EOM for varying values
of the radiative lifetime $T_1$ in the case of no insertion loss
(left y-axis) and the measured 1.9\,dB insertion loss (right
y-axis). For the measured $T_{1}\approx625$\,ps and
$T_{\text{mod}}$\,=\,370\,ps, the maximum and expected transmission
levels are 36\,$\%$ and 23\,$\%$, respectively.

Amplitude modulated QD emission is then sent to the HBT setup, and
$g^{(2)}(\tau)$ is measured as a function of excitation power, as
shown in Fig.~\ref{fig:fig2}d, with the unmodulated $g^{(2)}(\tau)$
measurements provided for reference.  A clear suppression in the
$g^{(2)}(0)$ values after modulation is observed, with improvements
ranging from a factor of eight at a pump power of
0.15$P_{\text{sat}}$ to a factor of two at 0.75$P_{\text{sat}}$. The
measured count rates on the Si SPADs after modulation are typically
$\approx$20\,$\%$ of the value before modulation.

The basic function of the modulator is to select a portion of the QD
emission with a user-defined width and center position.  Thus, if
the desired single photon emission has a different width and/or
temporal position with respect to multi-photon processes, the
amplitude modulation can discriminate between the two, removing the
undesired multi-photon emission.  To gain a better understanding of
how the timescale for single-photon and multi-photon emission differ
in this device, we measure $g^{(2)}(0)$ as a function of
$T_{\text{mod}}$ at an excitation power of 0.5$P_{\text{sat}}$, as
shown in Fig.~\ref{fig:fig2}e and in the Supplementary Information.
The nearly monotonic increase in $g^{(2)}(0)$ with increasing
modulation width shows that in this device, multi-photon emission is
spread over a timescale of a few ns.

The separation in timescales for single- and multi-photon emission
most likely depends on specific characteristics of the device in
question, including the pumping scheme and properties of the cavity
mode and its detuning with respect to the QD exciton state.
Recapture processes in the QD that lead to multiple photon emission
events from the QD excitonic line within a single excitation
pulse~\cite{ref:Peter_APL,ref:Santori_NJP,ref:Aichele_Zwiller_Benson}
represent one scenario in which such temporal separation may occur.
Alternately, recent
studies~\cite{ref:Chauvin_Fiore,ref:Laucht_Finely_2010} have
examined the differences in temporal behavior between single exciton
and multi-excitonic transitions of the QD, and have observed that
the emission processes can be delayed with respect to each other.

\begin{figure}[t]
\centering
\centerline{\includegraphics[width=8.5cm,clip=true]{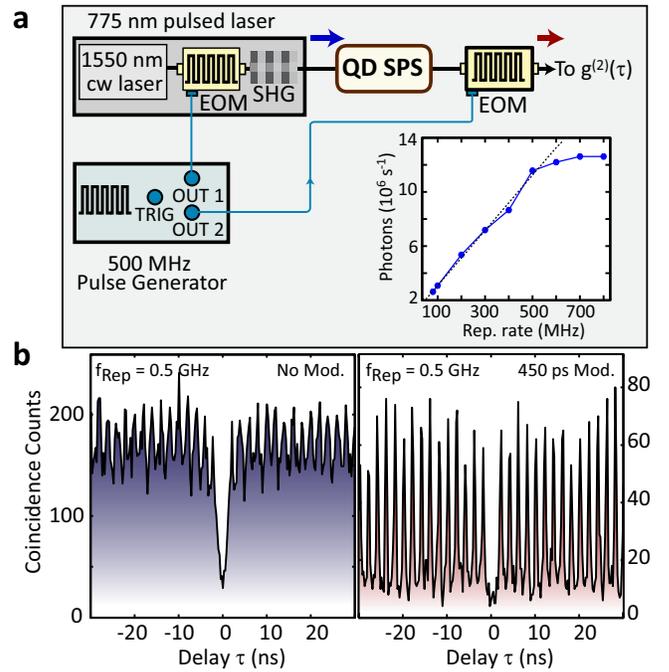}}
\caption{\textbf{Towards a GHz repetition rate QD SPS.} \textbf{a},
Setup for generating a 0.5\,GHz repetition rate QD SPS. The
excitation source is a modulated and frequency doubled 1550\,nm
laser, producing $\approx$250~ps width pulses at 775~nm. The pulse
generator driving the excitation source synchronously drives the
980\,nm EOM to modulate the QD emission. The bottom right inset
shows the collected photon count rate from the unmodulated QD SPS as
a function of the repetition rate. \textbf{b}, (left)
$g^{(2)}(\tau)$ without modulation. (right) $g^{(2)}(\tau)$ with
450\,ps modulation. A clear improvement in the overlap between
adjacent peaks in $g^{(2)}(\tau)$ is established after modulation.}
\label{fig:fig3}
\end{figure}

The temporal filtering provided by amplitude modulation can also be
useful in QD SPSs that operate at higher repetition rates. For a
source with pure single photon emission, the maximum repetition rate
depends on the radiative dynamics of the QD, including the carrier
capture time and QD radiative lifetime $T_{1}$. Purcell enhancement
to shorten $T_1$~[\onlinecite{ref:Ellis_NJP_fast_SPS}] and rapid
quenching of the QD emission at a timescale
$<T_1$~[\onlinecite{ref:Bennett_GHz_PRB}] have been used to approach
GHz repetition rates.  However, processes that lead to multi-photon
emission can be a limitation. Considering the aforementioned carrier
recapture processes, even if they still allow photons to be emitted
one at a time, multiple emission events per excitation cycle will
degrade the on-demand functionality of the source. Experimentally,
researchers have attributed various features in $g^{(2)}(\tau)$ data
to such processes~\cite{ref:Aichele_Zwiller_Benson,ref:Santori_NJP}.
For example, while measurements of QD SPSs under above-band
excitation do exhibit a pronounced antibunching dip (because photons
are emitted one-by-one), emission events that are asynchronous with
the excitation trigger lead to an overall background in
$g^{(2)}(\tau)$ at other times. Such behavior is exhibited in our
data without amplitude modulation (upper graphs in
Fig.~\ref{fig:fig2}d), and suggests that a higher repetition rate
source would benefit from suppression of events between the peaks.
The data taken after amplitude modulation (lower graphs
Fig.~\ref{fig:fig2}d) clearly shows such suppression, with
essentially no photon counts present in the regions between
successive peaks.

To demonstrate this technique in conjunction with a high repetition
rate QD SPS directly, we optically pump the device by an adjustable
repetition rate 775~nm source (Fig.~\ref{fig:fig3}a) whose trigger
output is synchronized to the 980\,nm EOM that modulates the
generated QD emission.  The limit on the useful repetition rate is
evident in the bottom right inset of Fig.~\ref{fig:fig3}a, which
shows how the collected QD emission level scales with repetition
rate, in the case of no amplitude modulation. For rates up to
0.5~GHz, the number of collected photons increases nearly linearly,
while faster repetition rates are precluded by the dynamics of the
QD. Figure~\ref{fig:fig3}b shows a measurement of $g^{(2)}(\tau)$
under this 0.5\,GHz excitation rate, without modulation (left) and
with $T_{\text{mod}}$\,=\,450\,ps$\pm$20\,ps (right).  The modulated
data displays a strong reduction in the overlap between peaks in
$g^{(2)}(\tau)$ (see Supplementary Information for additional data).
We also find that suppression of $g^{(2)}(\tau)$ in the regions
between the peaks does not necessarily require
$T_{\text{mod}}<T_{1}$; the Supplementary Information shows
$g^{(2)}(\tau)$ data in which the coincidences between peaks are
strongly suppressed even for $T_{\text{mod}}=$1.5~ns
$>T_{1}$=625~ps. In this scenario, amplitude modulation can be a
valuable resource in purifying and temporally separating the single
photon emission, with an overall transmission level that can be
$>60~\%$ (Fig.~\ref{fig:fig2}c). On the other hand, as we describe
in the following section, more aggressive amplitude modulation with
$T_{\text{mod}}<T_{1}$ can be used to improve the
indistinguishability of the source.

\vspace{3mm} \noindent \textbf{Improving the indistinguishability of
the single photon source}

The generation of indistinguishable photons is an important
requirement for several applications in quantum information
technology, such as linear optics quantum
computing~\cite{ref:Knill}, which relies on the two-photon
interference effect of single photon pulses at a beamsplitter. When
two indistinguishable photons enter a beamsplitter at the same time,
they bunch together and leave from the same exit
port~\cite{ref:Hong_Ou_Mandel_PRL}.  This can only be achieved if
the photon pulses are Fourier-transform limited, that is, the
coherence time ($T_2$) of the interfering photons is limited only by
their radiative lifetime ($T_1$), such that
$V\,=\,T_2$/(2$T_1$)\,=\,1. $V$ quantifies the degree of two-photon
wavpacket overlap, and in the limit of a pure SPS
($g^{(2)}(0)\,=\,0$), $V$\,=\,1 implies perfectly indistinguishable
photons.

\begin{figure}
\centering
\centerline{\includegraphics[width=8.5cm,clip=true]{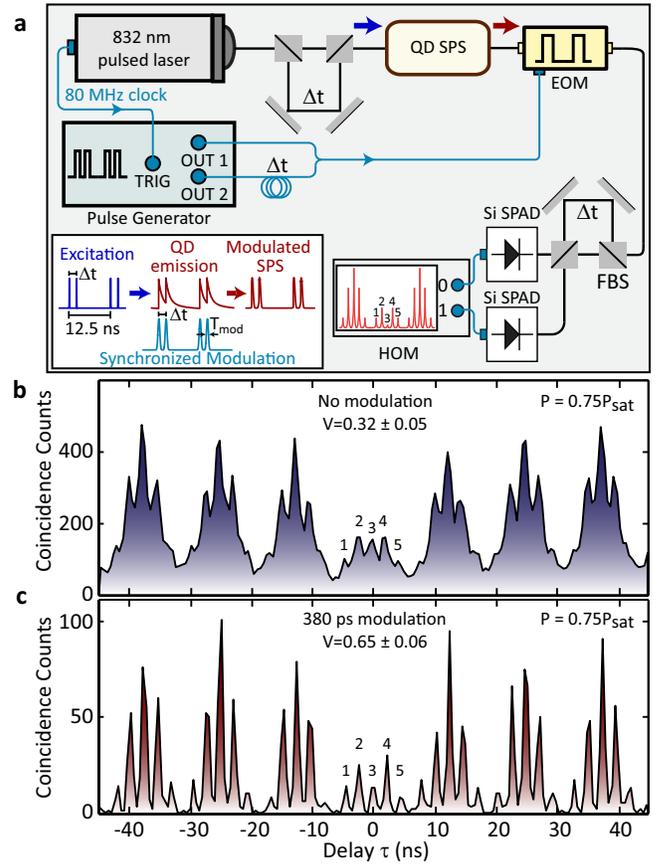}}
\caption{\textbf{Improving the indistinguishability of single
photons through amplitude modulation.} \textbf{a}, Schematic of the
setup for Hong-Ou-Mandel (HOM) interference with and without
amplitude modulation. Emission from the QD is sent into a
Mach-Zehnder interferometer in which a $\Delta$t\,=\,2.2\,ns delay
is inserted into one of the arms.  The same value of $\Delta$t is
used in the excitation path and in the dual-channel output of the
pulse pattern generator. \textbf{b}, HOM measurement without
amplitude modulation. \textbf{c}, HOM measurement with 380\,ps
amplitude modulation. The two-photon wavepacket overlap improves
from $V$\,=\,0.32$\,\pm\,$0.05 without amplitude modulation to
$V$\,=\,0.65\,$\pm$\,0.06 with amplitude modulation, in agreement
with the prediction based on the reduction from
$T_{1}\,=\,770$\,ps\,$\pm\,20$\,ps to
$T_{\text{mod}}\,=\,380$\,ps\,$\pm\,20$\,ps.  The uncertainties in
$V$ come from the uncertainty in the measured $g^{(2)}(0)$
values\cite{ref:QD_amplitude_modulation_note}.} \label{fig:fig4}
\end{figure}

The coherence time of single photons emitted from QDs is limited due
to several dephasing processes which reduce their
indistinguishability through lower $T_2$ values. Resonant
excitation~\cite{ref:Muller,ref:Vamivakas_Atature,ref:Ates_PRL09}
and Purcell enhancement can bring the photons closer to the
Fourier-transform limit~\cite{ref:Santori2,ref:Varoutsis_PRB05},
through the reduction of dephasing processes and the radiative
lifetime, respectively. Electrically-injected structures in which
dephasing was filtered out through fast Stark shifting have also
been demonstrated~\cite{ref:Bennett_indistinguishable_diode}. As a
new approach that is independent of the specific device geometry and
excitation wavelength, here we demonstrate that amplitude modulation
can improve the indistinguishability of our SPS through two means.
The first is through the improved purity of the SPS, as we have
detailed in the previous section. The second is through selection of
the coherent portion of the single photon wavepackets, which
increases $V$ from $T_{2}/(2T_{1})$ to $T_{2}/(2T_{\text{mod}})$.
Conceptually, this is similar to spectral filtering within the
homogeneous linewidth of the QD, which has been predicted to improve
photon indistinguishability~\cite{ref:Santori_NJP_Raman}.

Figure~\ref{fig:fig4}a shows the experimental setup used for photon
indistinguishability measurements. For each repetition period of the
832\,nm excitation laser, we generate a pair of pulses with a delay
$\Delta t\,=\,2.2$\,ns, equal to the delay in the HOM
interferometer, thus enabling the interference between the
consecutively emitted photons. The same delay is introduced to the
output of the electronic pulse generator which drives the EOM. For
these measurements, we use the same device as in the previous
section, but a shift in the spectral position of the cavity mode
with respect to the QD transition resulted in a longer radiative
lifetime $T_1$=$770$\,ps\,$\pm\,20$\,ps, and higher antibunching
value $g^{(2)}(0)\,=\,0.29\,\pm\,0.04$ (data in Supplementary
Information ~[\onlinecite{ref:QD_amplitude_modulation_note}]).
Figure~\ref{fig:fig4}b shows the result of the HOM measurement on
the spectrally filtered QD emission without amplitude modulation.
Examining the peak areas $A_{2,3,4}$ results in $M\,=\,0.49$, less
than the value $M\,=\,0.61$ expected for the measured $g^{(2)}(0)$
value if there was no two-photon
interference~\cite{ref:QD_amplitude_modulation_note}.

Next, we performed a HOM measurement after amplitude modulation. We
measured $T_{\text{mod}}\,=\,380$\,ps\,$\pm\,20$\,ps, which is
approximately half the QD $T_1$ value, and suggests that a factor of
two increase in $V$ should be expected.  We first measured the
auto-correlation of the modulated QD emission as
$g^{(2)}(0)\,=\,0.20\,\pm\,0.04$~\cite{ref:QD_amplitude_modulation_note},
again evidencing an improvement in the purity of the SPS.
Figure~\ref{fig:fig4}c shows the result of the HOM experiment, where
the correlation peaks are now well-separated due to the modulation.
We estimate $M\,=\,0.31$, which is smaller than the value
$M\,=\,0.58$ expected for this device if there was no two-photon
interference. Our measured value $M\,=\,0.31$ yields
$V$\,=\,0.65$\,\pm\,0.06$, which corresponds to a coherence time
$T_2\,=\,500$\,ps\,$\pm\,50$\,ps given
$T_{\text{mod}}\approx380$\,ps. In comparison, the unmodulated case
has $M\,=\,0.49$ and $g^{(2)}(0)\,=\,0.29\,\pm\,0.04$, which gives
an unmodulated value $V\,=\,0.32\,\pm\,0.05$ that is consistent with
the ratio $T_{2}/2T_{1}$ for $T_{2}\approx$500\,ps and
$T_{1}\approx$770\,ps. Thus, the two-photon wavepacket overlap $V$
is increased by a factor of two, as expected based on the change
from $T_1$ to $T_{\text{mod}}$ produced by amplitude modulation. $V$
after modulation approaches the value of $\approx$\,0.8 achieved in
previous
works~\cite{ref:Santori2,ref:Varoutsis_PRB05,ref:Weiler_Michler_PSS_B}
through quasi-resonant excitation and larger Purcell enhancement.
Amplitude modulation is fully compatible with such techniques, where
shorter $T_1$ values would result in higher transmission for a fixed
$T_{\text{mod}}$ (Fig.~\ref{fig:fig2}c), and longer $T_2$ values
would improve the indistinguishability.

\vspace{3mm} \noindent \textbf{Conclusions}

In conclusion, we have demonstrated that synchronous amplitude
modulation of QD emission can be an effective approach to improving
its performance as a SPS. The temporal filtering provided allows us
to select the portions of the emission for which the behavior is
more ideal. This results in a significant improvement in the single
photon purity of the source, by as much as a factor of 8, and
enables clean operation of the SPS up to repetition rates as high as
0.5\,GHz. Using amplitude modulation to eliminate portions of the
single photon wavepackets that are incoherent improves the
two-photon wavepacket overlap by a factor of 2.  Furthermore, the
ability to convert exponentially-shaped single photon wavepackets to
Gaussian-shaped ones may improve the robustness of the source in
some applications~\cite{ref:Rohde_PRA}.

Finally, we emphasize the versatility of amplitude modulation, as it
can be applied to either optically (non-resonant or resonant) or
electrically pumped devices, independent of device geometry and the
precise energy level structure of the QD.  It can also be used in
conjunction with other methods that improve the performance of SPSs,
such as resonant excitation or Purcell enhancement.  Other
solid-state quantum emitters, such as nitrogen vacancy centers in
diamond~\cite{ref:Kurtsiefer,ref:Babinec_Loncar_NV_SPS} and
colloidal quantum dots~\cite{ref:Sebald_Michler_CdSe_SPS} often
exhibit non-zero multi-photon probability and imperfect two-photon
interference; amplitude modulation may be a valuable resource for
those systems as well.  Finally, the ability to temporally filter
the emitted signal with adjustable width and position provides a new
resource to help understand the dynamics within mesoscopic quantum
systems like single semiconductor quantum dots.

\noindent \textbf{Acknowledgements} S.A. and I.A. acknowledge
support under the Cooperative Research Agreement between the
University of Maryland and NIST-CNST, Award 70NANB10H193. The
authors thank Matthew Rakher for useful discussions and early
contributions to this work.

\bibliographystyle{apsrev4-1}
\bibliography{KS_bib_2012_7_05}

\onecolumngrid
\newpage

\section{Supplementary Information}

\section{Experimental Details}

\subsection{Device Fabrication}

Devices are fabricated in a wafer grown by molecular beam epitaxy
and consisting of a single layer of InAs QDs embedded in a 190\,nm
thick layer of GaAs, which in turn is grown on top of a 1\,$\mu$m
thick layer of Al$_x$Ga$_{1-x}$As with an average $x$\,=\,0.65.  The
s-shell peak of the QD ensemble is located at 965\,nm, and a
gradient in the QD density is grown along one axis of the wafer.
Low-temperature photoluminescence measurements of the wafer are
performed prior to device definition to determine the appropriate
location on the wafer (in terms of QD density) to fabricate devices.

Microdisk cavities of varying diameter between 2\,$\mu$m and
4\,$\mu$m are fabricated through: (i) electron beam lithography,
(ii) resist reflow, (iii) Ar-Cl$_2$ inductively-coupled plasma
reactive ion etching of the GaAs layer and removal of the electron
beam resist, and (iv) (NH$_4$)$_2$S and HF wet etching of the
underlying Al$_x$Ga$_{1-x}$As layer to form the supporting pedestal.

\subsection{Si single-photon avalanche diode (SPADs)}

Two different types of Si single-photon avalanche diodes (SPADs) are
used in the experiments, depending on the requirements on detection
efficiency and timing resolution.  Thick Si SPADs used in this work
have a detection efficiency of $\approx$12.5\,$\%$ at 980\,nm and a
timing jitter $\approx$700~ps, and are used in experiments in which
faster timing resolution are not needed.  This includes the data for
Figures 1e, 2d, 4b, and 4c in the main text.

Newly-developed red-enhanced thin Si
SPADs\,\cite{ref:Gulinatti_R_SPAD} have a detection efficiency of
$\approx$6\,$\%$ at 980\,nm, and a timing jitter of
$\approx$100\,ps. They are used in experiments for which faster
timing resolution is needed, including Figs. 1c, 2b, and Fig. 3 in
the main text, and Fig.~\ref{fig:SFig4} and \ref{fig:SFig5} in the
Supplementary Material.  The outputs of the Si SPADs are fed to a
time-correlated single photon counting (TCSPC) board for all photon
correlation measurements.

\subsection{Spectral filtering setup}

QD emission that is out-coupled into the FTW is sent into a
$\approx$\,0.2\,nm bandwidth volume reflective Bragg grating whose
input is coupled to single mode optical fiber and output is coupled
to polarization maintaining (PM) single mode fiber. Quarter- and
half-wave plates and a polarizing beamsplitter are placed prior to
the PM fiber, to ensure that light is linearly polarized along the
slow-axis of the fiber. The typical throughput of the filtering
setup is $\approx$50$\,\%$.

\subsection{Electro-optic modulator setup}

A 980\,nm band, fiber-coupled LiNbO$_3$ electro-optic modulator
(EOM) is used for all amplitude modulation experiments.  Both the
modulator input and output are PM fibers, with the polarization
aligned along the slow-axis of the fiber.  The modulator is driven
by a dual-channel electronic pulse generator that can be internally
or externally triggered and can produce pulses as short as 250\,ps.
For the experiments in Fig.~2 and Fig.~3b in the main text, one
channel of the pulse generator is used to generate a periodic train
of pulses which drive the EOM. For the experiments in Fig.~4c and
Fig.~\ref{fig:SFig4}b, a pair of pulses per repetition period are
generated by combining both output channels of the generator, with
an electronically adjustable delay between them. The pulse generator
output (either single channel or combined double channel) is fed to
a $12.5$\,GHz amplifier to generate the required voltage needed to
achieve maximum transmission through the EOM. A separate DC power
supply is used to control the bias voltage on the EOM in all
experiments.

While the electronic pulse generator produces output pulses as short
as 250\,ps, addition with the second output channel (for double
pulse experiments), amplification, and application to the optical
signal results in optical pulses that are typically broader.  The
narrowest pulse widths achieved, as measured on the TCSPC, are
$\approx$350\,ps.

\subsection{Lifetime measurements}

Measurements of the QD lifetime are performed by sending the trigger
output of the 832\,nm excitation laser to the start input of the
TCSPC, and filtered QD emission to a fast Si SPAD whose output is
sent to the stop input of the TCSPC.  The bin size of the TCSPC is
typically set to 32\,ps, near the timing jitter of the SPAD and well
below the lifetimes measured in this work.

\subsection{HBT and HOM setups}

\begin{figure}[h]
\centerline{\includegraphics[width=0.8\linewidth]{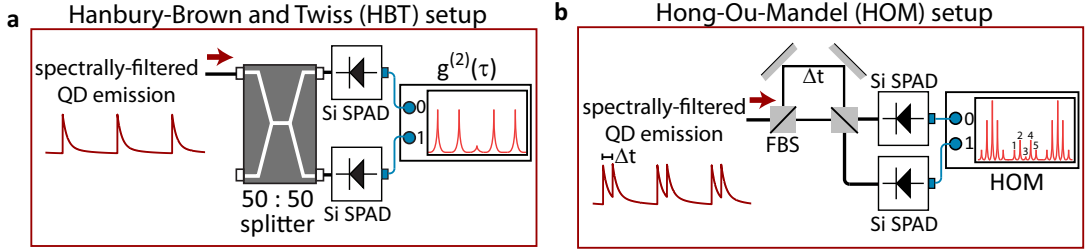}}
\caption{Setups for \textbf{a}, Hanbury-Brown and Twiss (HBT) photon
correlation measurement, and \textbf{b}, Hong-Ou-Mandel (HOM) photon
indistinguishability measurement.  FBS=fiber-coupled beasmplitter,
SPAD=single photon avalanche diode.} \label{fig:SFig1}
\end{figure}

The second-order correlation function $g^{(2)}(\tau)$ is measured by
using a Hanbury-Brown and Twiss setup, which consists of a 1x2 fiber
splitter and two Si SPADs, as shown in Fig.~\ref{fig:SFig1}a. The
spectrally filtered QD emission is connected to the input port of
the splitter and the outputs of the splitter are connected to the
SPADs. The output of the SPADs is fed to the TCSPC module.  For
Figs.~1 and 2 in the main text, and Fig.~\ref{fig:SFig3} in the
Supplementary Material, thick Si SPADs are used and data from the
TCSPC is acquired in histogram mode with a bin size of 512\,\,ps.
For experiments in Fig.~3 in the main text and Fig.~\ref{fig:SFig4}
and \ref{fig:SFig5} in the Supplementary Material, red-enhanced thin
Si SPADs are used, and data is acquired in a time-tagged,
time-resolved mode in which the photon arrival times from each
channel are recorded with 4\,ps timing resolution and the bin size
set during subsequent data analysis. A bin size of 256\,ps was set
for Fig.~3, while a bin size of 200\,ps was set for
Fig.~\ref{fig:SFig4}.

Indistinguishability of single photon pulses is measured by using a
PM fiber-based Hong-Ou-Mandel (HOM) type interferometer (See
Fig.~\ref{fig:SFig1}b). The experiments rely on two-photon
interference between consecutively emitted single
photons~\cite{ref:Santori2}. For this purpose, each excitation pulse
(with a 80 MHz repetition rate) is split to create two pulses with a
delay $\Delta$t\,=\,2.2\,ns, which then generate a pair of single
photon pulses from the QD transition. As described above, the
spectral filtering setup leaves the emission linearly polarized,
after which it is coupled to the HOM setup, which uses two fiber
optic non-polarizing beamsplitters (FBS) with a fixed delay
$\Delta$t\,=\,2.2\,ns between the interferometer arms. Two-photon
interference occur at the second FBS, the output ports of which are
connected to thick Si SPADs for detection of photon coincidences.
The outputs of the two SPADs are sent to the TCSPC, which is
operated in histogram mode with a bin size of 512\,ps.

Figure~\ref{fig:SFig1}b shows a typical correlation histogram, which
has five-peak clusters with a spacing defined by the repetition rate
of the excitation laser (12.5~ns). The five peaks within each
cluster are separated by 2.2\,ns, given by the delay introduced
between the interferometer arms. The two-photon interference effect
is reflected in the central cluster, where the peaks named from 1 to
5 are formed due to different combinations of paths taken by the
interfering photons.The outer side peaks 1 and 5 correspond to the
case where the first photon follows the short arm and the second
photon follows the long arm, while the inner side peaks 2 and 4
arise if both photons follow the same arm of the interferometer.
Finally, peak 3 at $\tau$\,=\,0 corresponds to the case where the
first photon follows the long arm and the second photon follows the
short arm, which leads to the overlap and interference of
consecutively emitted photons at the second BS.

\section{Expected transmission through the Electro-Optic Modulator}

We quantitatively describe the fraction of photons transmitted by
the EOM using the approach in Ref.~\onlinecite{ref:Rakher_APL_2011}.
We assume that the modulation has a Gaussian temporal profile given
by
\begin{equation}
M(t) = \exp\left[ \frac{-t^2}{\sigma^2} \right],
\end{equation}
\noindent where $T_{\text{mod}}$=$2\sigma$ is the full-width of the
modulation pulse at its 1/e point. We further take the probability
distribution of photons emitted by the QD to be given by
\begin{equation}
R(t) = \frac{1}{T_1}H(t)\exp(-t/T_1),
\end{equation}
\noindent where $H(t)$ is the Heaviside step function.  The fraction
of transmitted photons is then
\begin{equation}
f(\Delta T_{mod}) = \int \textrm{d}t R(t) M(t-\Delta T_{mod}),
\end{equation}
\noindent where $\Delta T_{mod}$ is the delay of the modulation gate
with respect to the incident QD photon.  Maximizing $f$ with respect
to $\Delta T_{mod}$ gives a plot of the transmission as a function
of $T_{\text{mod}}$ for a given value of $T_{1}$, as shown in
Fig.~2c in the main text. The right y-axis in that plot shows the
expected transmission if an additional 1.9\,dB of insertion loss
(the measured value for the modulator used in our experiments) is
included.

\section{Supplementary Measurements and Data Analysis}

\subsection{SPS Brightness Estimate} The QD SPS efficiency is
estimated by comparing the number of photons coupled to the FTW at
$P_{\text{sat}}$ to the expected number of generated photons. The
total detection efficiency of the optical setup $\zeta$ includes the
transmission through the FTW (50\,\%) and spectral filtering setup
(50\,\%) and the quantum efficiency of the SPADs (12.5\,\%),
measured using a laser of known power at the QD emission wavelength.
Assuming the QD generates one photon per excitation pulse, the
efficiency of the QD SPS is given by
$\xi\,=\,I_{sat}/(R_{Rep}*\zeta)$, where $I_{sat}$ is the detected
count rate on the SPAD and $R_{Rep}$ is the 80\,MHz repetition rate
of the excitation laser. Error bars in the measurements come from
the fluctuation in the detected count rates, and are one standard
deviation values.

\subsection{Cavity transmission measurement}

Measurement of the cavity transmission spectrum is done sweeping the
wavelength of a 980\,nm band external cavity tunable diode laser
that is connected to the FTW input and measuring the transmitted
intensity on an InGaAs photoreceiver that is connected to the FTW
output.

The transmitted signal on resonance can be written
as~\cite{ref:Barclay7}

\begin{equation}
T = \frac{(1-K)^2}{(1+K)^2}
\end{equation}

\noindent where $K$ is a coupling parameter given by the ratio of
the FTW-cavity mode coupling rate to all other losses (parasitic and
intrinsic) in the system.  The fraction of cavity mode photons that
are out-coupled into the FTW is then given by:

\begin{equation}
\eta=\frac{1}{1+1/K}
\end{equation}

Ideal microdisk cavities support degenerate clockwise and
counterclockwise whispering gallery modes (WGMs).  While this
degeneracy can be broken in the limit of strong backscattering, in
this work the backscattering rate is much smaller than the intrinsic
loss rate, as evidenced by the single dip present in the
transmission spectrum. The QD, however, will emit into both WGMs,
with the clockwise WGM coupling to the forward channel of the FTW,
and the counterclockwise WGM coupling to the backward channel of the
FTW. Based on the transmission spectrum in Fig.~1c of the main text,
$K\approx0.33$, resulting in $\eta\approx$25\,$\%$ for each channel
of the FTW.

\subsection{Micro-photoluminescence measurements}

\begin{figure}[h]
\centerline{\includegraphics[width=0.6\linewidth]{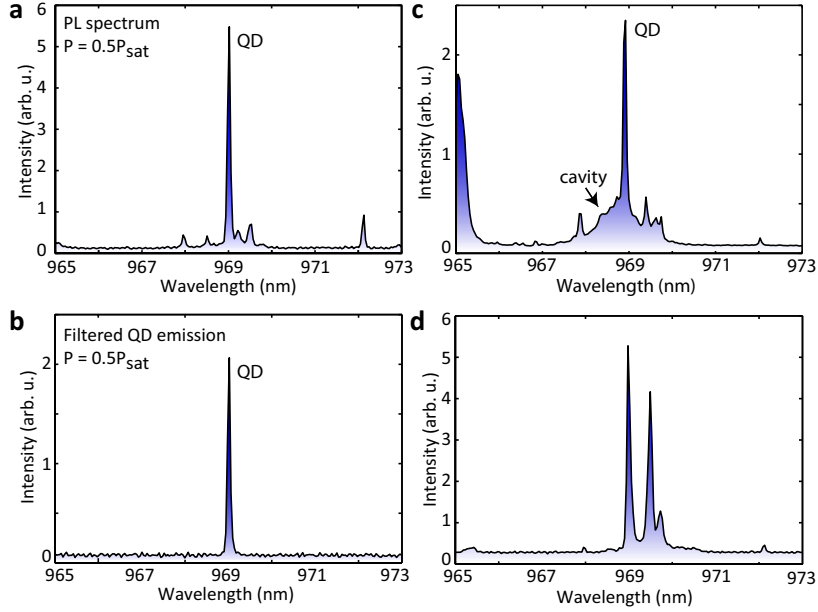}}
\caption{Micro-PL data, showing \textbf{a}, a PL spectrum near
saturation, where a single QD excitonic line is enhanced by a broad
cavity mode, \textbf{b}, the spectrally filtered QD line that is
sent into HBT and HOM setups, \textbf{c}, the PL spectrum far above
saturation, in order to demonstrate the location of the cavity mode
and broad multi-excitonic emission, and \textbf{d}, the PL spectrum
under a different cavity detuning and FTW-cavity coupling condition,
where multiple bright excitonic states are observed.}
\label{fig:SFig2}
\end{figure}

Figure~\ref{fig:SFig2} shows micro-photoluminescence data from the
microdisk-QD device studied in the main text under a variety of
conditions. Spectra taken before and after spectral filtering, and
under saturation conditions with 832\,nm pulsed pumping, are shown
in Fig.~\ref{fig:SFig2}a-b.  The QD sits on a broad cavity mode
which provides both Purcell enhancement of the QD radiative rate and
efficient out-coupling into the FTW.  The location of the cavity
mode can be seen under strong excitation conditions, well beyond the
QD saturation power (Fig.~\ref{fig:SFig2}c); transmission
measurements as discussed above and provided in Fig.~1c in the main
text are also used.  Finally, adjustment of the cavity mode spectral
position can enable multiple QD excitonic lines to be collected with
high efficiency, as seen in Fig.~\ref{fig:SFig2}d.

\subsection{Lifetime measurements}

Lifetime data is fit to an exponential and the quoted uncertainties
are the one-standard deviation values. Modulated lifetimes are
measured in the same way, where now the QD emission has first gone
through the EOM before being detected by the SPAD. Quoted modulation
widths are given by the full-width at the 1/e points of the data.

\subsection{$g^{(2)}(\tau)$ measurements}

Measurement of the second-order correlation function $g^{(2)}(\tau)$
are performed using the setups described above. In pulsed
measurements, the $g^{(2)}(0)$ value is determined by comparing the
integrated area of the peak around time zero to the average area of
the peaks away from time zero. The uncertainty on this value is
given by the standard deviation in the area of the peaks away from
time zero.

\subsection{Modulation-width dependent measurements}

\begin{figure}[h]
\centerline{\includegraphics[width=0.45\linewidth]{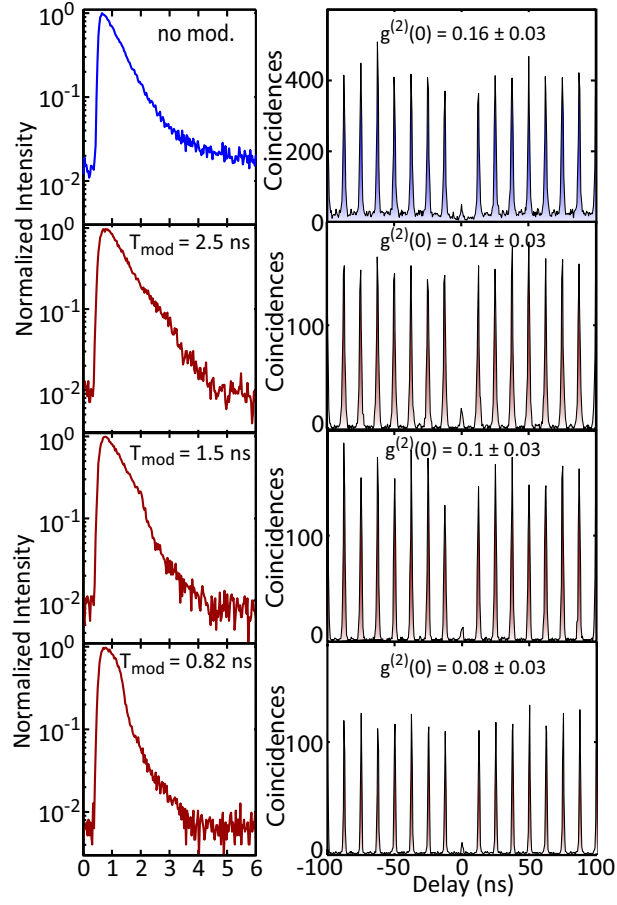}}
\caption{Modulation-width dependent lifetime and $g^{(2)}(\tau)$
measurements, for the device studied in the main text in Fig.~2.}
\label{fig:SFig3}
\end{figure}

Adjusting the modulation width is expected to influence qualities
like the QD SPS purity and indistinguishability.  Here, we provide
the individual $g^{(2)}(\tau)$ measurements used to produced
Fig.~2e, which plotted $g^{(2)}(0)$ as a function of
$T_{\text{mod}}$. Figure~\ref{fig:SFig3}a shows a few measured
lifetimes and $g^{(2)}(\tau)$ curves for $T_{\text{mod}}$ between
820\,ps and 2500\,ps, with the unmodulated case from the main text
repeated for reference. In comparison to the extracted $g^{(2)}(0)$
values, the full $g^{(2)}(\tau)$ curves present additional
information, indicating, for example, the values of $T_{\text{mod}}$
for which the overlap between adjacent peaks is suppressed.  Here,
we see the possible benefits of intermediate modulation widths; at
$T_{\text{mod}}$\,=\,820\,ps, the overlap is almost completely
suppressed, $g^{(2)}(0)\,=\,0.08\,\pm\,0.03$ is reduced by a factor
of almost 2, and the transmission through the setup (including
insertion loss) is $\approx$37\,$\%$, nearly a factor of two higher
than in the 370\,ps modulation width case.

\subsection{Temporal overlap and high repetition rate sources}

Figure~3 in the main text shows $g^{(2)}(\tau)$ for the microdisk-QD
device under 0.5~GHz excitation rate, with and without synchronous
amplitude modulation of the QD emission.  While modulation clearly
improves the temporal separation between the peaks, the degree to
which this is improved is somewhat less than that observed under
80~MHz pumping in Fig.~2 in the main text, and Fig.~\ref{fig:SFig3}
above. We believe that this is primarily due to the characteristics
of the source exciting the QD.  Because the excitation source was
produced by modulating and frequency doubling a cw 1550\,nm laser,
both the excitation wavelength (775\,nm) and the pulse width
($\approx250$\,ps) differ from the mode-locked laser source used in
the rest of the work (wavelength of 832\,nm and pulse width
$<10$\,ps).  The significantly increased pulse width, in particular,
begins to approach the radiative lifetime of the QD and may lead to
much stronger carrier recapture and multi-photon emission processes.

\begin{figure}[h]
\centerline{\includegraphics[width=0.75\linewidth]{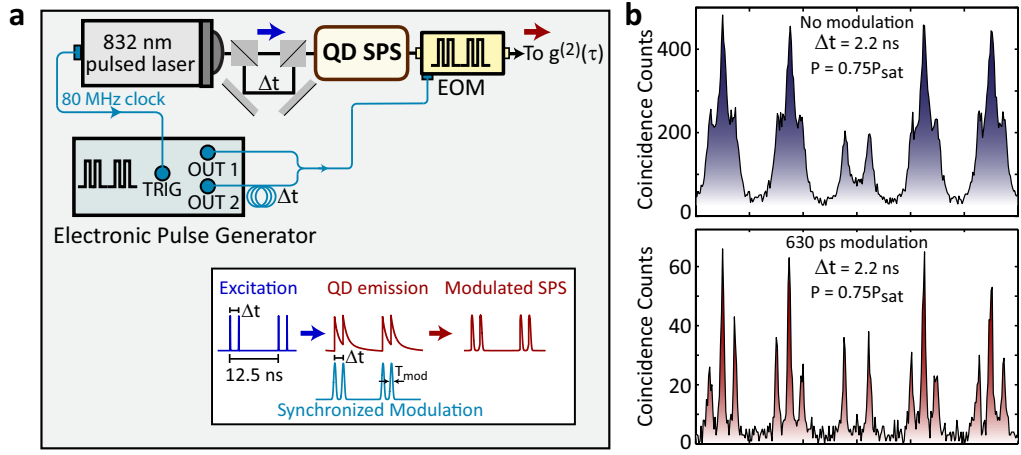}}
\caption{\textbf{a}, Setup for checking the temporal overlap of
successive photon emission events. A variable delay produces a pair
of excitation pulses separated by $\Delta$t for each repetition
period of 12.5\,ns. The pulse generator produces a pair of
electronic pulses, with the same delay, that drive the EOM situated
at the output of the SPS. \textbf{b}, Top panel: $g^{(2)}(\tau)$
under double pulse excitation and $\Delta$t\,=\,2.2\,ns without
modulation, Bottom panel: $g^{(2)}(\tau)$ under double pulse
excitation and $\Delta$t\,=\,2.2\,ns with 630\,ps modulation. }
\label{fig:SFig4}
\end{figure}

To gain an understanding of how the system might respond under a
high repetition rate but with a short excitation pulse width, we use
the setup shown in Fig.~\ref{fig:SFig4}a, where a pair of
beamsplitters and mirrors have been placed in the 832\,nm excitation
path to generate two pump pulses for every excitation period, with a
delay between the pulses $\Delta t$ that is adjustable.  The output
of the QD SPS is hooked up to an EOM that is driven by a pulse
generator that creates a pair of pulses with adjustable modulation
width $T_{\text{mod}}$ and a delay between the pulses that matches
the excitation delay $\Delta t$.

The top panel of Fig.~\ref{fig:SFig4}b shows a measurement of
$g^{(2)}(\tau)$ for $\Delta t$\,=\,2.2\,ns, without amplitude
modulation. Ideally, one would expect a series of three-peak
clusters separated by the repetition period (12.5\,ns), where the
peaks within the cluster are separated by $\Delta t$, and the
central peak at $\tau$\,=\,0 vanishes for a pure single photon
source.  We qualitatively observe these features, but there is a
strong overlap of the peaks, suggesting that a further reduction in
$\Delta t$ would not be feasible.  In comparison, modulation with
$T_{\text{mod}}$\,=\,630\,ps$\,\pm\,20$\,ps results in the bottom
panel of Fig.~\ref{fig:SFig4}b, where the overlap has been
significantly diminished, nearing the levels seen in the 80~MHz
repetition rate measurements.  This suggests that the 0.5~GHz
repetition rate measurements of Fig.~3b in the main text would show
additional improvement in the background coincidence levels between
the peaks under appropriate short-pulse excitation.

\subsection{Photon indistinguishability measurements}

Indistinguishability of single photon pulses is measured by using
the two-photon interference effect on a beamsplitter. The degree of
indistinguishability is mainly given by the mean overlap of the
wavepackets of the interfering photons. Perfect overlap will only
occur if the interfering photons are Fourier-transform limited,
which requires that the coherence time $T_2$ of the photons is
mainly limited by the radiative lifetime $T_1$ ($T_2$\,=\,2$T_1$).
The degree of indistinguishability is therefore quantified by the
two-photon overlap $V$ and given by~\cite{ref:Santori2}:

\begin{IEEEeqnarray}{rCl}
    V = \frac{T_2}{2T_1} = \frac{\gamma}{\gamma + \alpha}
\end{IEEEeqnarray}

\noindent where, $\gamma$ is the spontaneous emission amplitude
decay rate ($\gamma\,=\,1/2T_1$) and $\alpha$ is the pure dephasing
rate ($\alpha\,=\,1/T^{\star}_2$, $T^{\star}_2$ is the pure
dephasing time), which is linked to the coherence time $T_2$ via:

\begin{IEEEeqnarray}{rCl}
    \frac{1}{T_2} = \frac{1}{2T_1} + \frac{1}{T^{\star}_2}
\end{IEEEeqnarray}

Experimentally, the degree of indistinguishability is quantified
from the mean areas of the correlation peaks named from 1 to 5 in
Fig.~\ref{fig:SFig1}b, which are calculated as~\cite{ref:Santori2}:

\begin{IEEEeqnarray}{rCl}
    A_1 & = & N\eta^{(2)}R^{3}T \nonumber \\
    A_2 & = & N\eta^{(2)}[R^{3}T(1+2g^{\star})+RT^{3}] \nonumber\\
    A_3 & = & N\eta^{(2)}[(R^{3}T+RT^{3})(1+2g^{\star})-2(1-\epsilon)^{2}R^{2}T^{2}V(\Delta t)]\\
    A_4 & = & N\eta^{(2)}[R^{3}T+RT^{3}(1+2g^{\star})] \nonumber\\
    A_5 & = & N\eta^{(2)}R^{3}T\nonumber
\end{IEEEeqnarray}

\noindent where $N$ is the number of repetitions, $\eta^{(2)}$ is
the combined two-photon generation and detection efficiency, $R$ and
$T$ are the reflection and transmission coefficients of the
beamsplitters, (1-$\epsilon$) is the visibility of the interference
setup, $g^{\star}$ is the two-photon emission probability, and $V$
is the mean two-photon overlap. The effect of two-photon
interference is reflected as a reduced area of peak 3 and it is
quantified by using the areas of Peaks 2, 3 , and 4:

\begin{IEEEeqnarray}{rCl}
    M = \frac{A_3}{A_2+A_4} =  \frac{(1+2g^{\star})}{2(1+g^{\star})}-\frac{(1-\epsilon)^{2}R^{2}T^{2}V}{(1+g^{\star})(R^{3}T+RT^{3})}
\label{eq:M_eqn}
\end{IEEEeqnarray}

The $M$ parameter defines the probability of two photons merging in
the beamsplitter and leaving in opposite directions. Theoretically,
the value of $M$ will be between 0 and 0.5, depending on the value
two-photon overlap, in the case of perfect visibility of the
interferometer setup ((1-$\epsilon$)\,=\,1), a perfect 50/50
splitting ratio of the splitters ($R\,=\,T\,=\,0.5$), and
$g^{\star}\,=\,0$.

\begin{figure}[h]
\centerline{\includegraphics[width=0.7\linewidth]{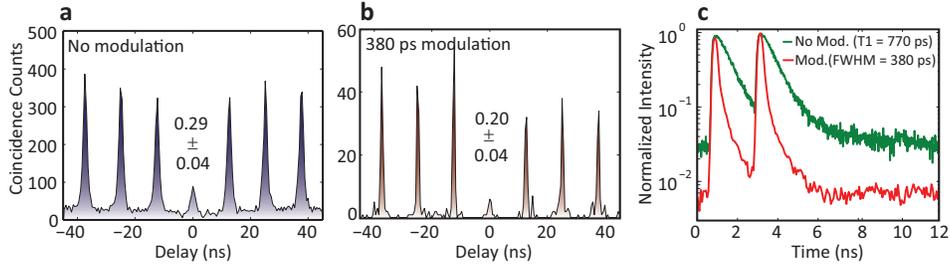}}
\caption{Supporting data for the Hong-Ou-Mandel measurements
presented in the main text in Fig.~4. \textbf{a}, $g^{(2)}(\tau)$
data measured without amplitude modulation, showing
$g^{(2)}(0)\,=\,0.29\,\pm\,0.04$. \textbf{b}, $g^{(2)}(\tau)$ data
measured with 380\,ps modulation, showing
$g^{(2)}(0)\,=\,0.20\,\pm\,0.04$. \textbf{c}, Double-pulse lifetime
without modulation shown in green, with an extracted
$T_{1}$\,=\,770\,ps. Double-pulse lifetime after modulation shown in
red, with an extracted $T_{\text{mod}}$\,=\,380\,ps.}
\label{fig:SFig5}
\end{figure}

Now, we discuss the details of the experimental results using the
expressions given above. We assume (1-$\epsilon$)\,=\,1 and
$R\,=\,T\,=\,0.5$ for all data. Figure~\ref{fig:SFig5}a shows the
result of photon correlation measurement performed on the QD
emission before amplitude modulation under the same conditions as
the HOM measurement shown in the main text (See Fig.~4b). We measure
$g^{(2)}(0)\,=\,0.29\,\pm\,0.04$, which places an upper limit (based
on Eqn.~\ref{eq:M_eqn}) for $M\,=\,0.61$ in the case of no
interference ($V\,=\,0$). Here we use $g^{(2)}(0)\,=\,g^{\star}$
since we didn't observe any long time blinking
behavior~\cite{ref:Santori2}. By using the ratio of the areas from
the experimental HOM data, we estimate the value of
$M\,=\,0.49\,<\,0.61$, indicating a two-photon overlap
$V\,=\,0.32\,\pm\,0.05$ as a result of a partial
indistinguishability. This number is consistent with that expected
from the ratio $T_2/(2T_1)$, with $T_2$\,=\,500\,ps and
$T_1$\,=\,770\,ps (Fig.~\ref{fig:SFig5}c).

Similar measurements were performed after amplitude modulation,
where $T_{\text{mod}}=380$\,ps is measured (Fig.~\ref{fig:SFig5}c)
to be approximately half of the emission lifetime $T_1\,=\,770$\,ps.
Figure~\ref{fig:SFig5}b shows the result of a photon correlation
measurement performed on the QD emission under the same conditions
as the HOM measurement shown in the main text (See Fig.~4c). An
antibunching value $g^{(2)}(0)\,=\,0.20\,\pm\,0.04$ is measured,
which places a values $M\,=\,0.58$ in the limit of no two-photon
interference ($V\,=\,0$). We estimated the $M$ value from the
experimental HOM data as $M\,=\,0.31$, which corresponds to a
two-photon overlap $V\,=\,0.65\,\pm\,0.06$. This value is almost a
factor two higher than the one obtained before amplitude modulation,
as expected since $T_{\text{mod}}\,\approx\,0.5T_1$.

HOM measurements performed in Fig.~1e yielded $M\,=\,0.40$ for
$g^{(2)}(0)\,=\,0.16\,\pm\,0.01$, smaller than the value predicted
in the case of no two-photon interference ($V\,=\,0$), $M\,=\,0.57$.
The corresponding two-photon overlap $V\,=\,0.39\,\pm\,0.05$
determined from Eqn.~\ref{eq:M_eqn} matches well with the prediction
$T_2/(2T_1)$, with $T_2$\,=\,500\,ps and $T_1$\,=\,625\,ps.

\end{document}